\documentclass[11pt,a4paper]{article}

\usepackage[T1]{fontenc}
\usepackage[utf8]{inputenc}
\usepackage{lmodern}
\usepackage[margin=1.5in]{geometry}
\usepackage[english]{babel}
\usepackage{amsmath,amssymb}
\usepackage[numbers,sort&compress]{natbib}
\usepackage[hidelinks]{hyperref}
\usepackage{microtype}

\title{Revised comment on the paper titled ``The Origin of Quantum Mechanical Statistics: Insights from Research on Human Language''\\
(arXiv preprint arXiv:2407.14924, 2024)}

\author{Mikołaj Sienicki\thanks{Polish-Japanese Academy of Information Technology, ul. Koszykowa 86, 02-008 Warsaw, Poland, European Union.} \and
Krzysztof Sienicki\thanks{Chair of Theoretical Physics of Naturally Intelligent Systems (NIS), Lipowa 2/Topolowa 19, 05-807 Podkowa Leśna, Poland, European Union.}
}

\date{\today}

\begin{document}

\hyphenpenalty=10000
\exhyphenpenalty=10000
\sloppy

\maketitle

\begin{abstract}
This short note comments on \citet{Aerts2024Origin}, which proposes that ranked word frequencies in texts should be read through the lens of Bose--Einstein (BE) statistics and even used to illuminate the origin of quantum statistics in physics. The core message here is modest: the paper offers an interesting analogy and an eye-catching fit, but several key steps mix physical claims with definitions and curve-fitting choices. We highlight three such points: (i) a normalization issue that is presented as ``bosonic enhancement,'' (ii) an identification of rank with energy that makes the BE fit only weakly diagnostic of an underlying mechanism, and (iii) a baseline comparison that is too weak to support an ontological conclusion. We also briefly flag a few additional concerns (interpretation drift, parameter semantics, and reproducibility).
\end{abstract}

\section{What the paper claims, in plain terms}
\citet{Aerts2024Origin} propose a mapping from a text to an ``ideal gas'' picture: word-types are treated as if they were particles occupying ``energy levels,'' where the level index is simply the word's rank in the frequency table. A Bose--Einstein-shaped occupancy curve is then fitted to the rank--frequency list, and the quality of the fit is taken to support a stronger interpretation---that texts behave like a gas of indistinguishable bosons, and that this analogy may even shed light on why Bose--Einstein statistics appears in physics.

There is nothing wrong with exploratory analogies. The issue is that the paper repeatedly slides from ``this curve fits'' to ``this is evidence for a specific physical mechanism.'' The three points below explain why that slide is not justified by the present analysis.

\section{Three core technical concerns}

\subsection{Normalization does not create a probability boost}
A central step argues that when two single-particle states are set equal inside a symmetrized two-boson expression, the state vector acquires a factor $\sqrt{2}$ and hence the squared norm becomes $2$, which is then read as a doubling of the probability that two bosons occupy the same microstate \citep{Aerts2024Origin}.
But an overall scale factor of a ket is not a physical probability. Probabilities are computed from \emph{normalized} states; rescaling a vector does not change physics. To be clear: (anti-)symmetrization \emph{can} change joint detection statistics once an observable and a measurement scenario are specified, but the mistake is to read the norm of an \emph{unnormalized} ket as a propensity. If the intended point is bosonic ``bunching,'' that phenomenon arises from interference in a \emph{specified measurement set-up} (e.g.\ Hong--Ou--Mandel-type effects), not from treating the norm of an unnormalized ket as a probability \citep{HOM1987}.

\subsection{Rank-as-energy makes the BE fit only weakly diagnostic}
The ``energy levels'' used in the paper are defined by rank,
\begin{equation}
E_i = i,
\label{eq:rankenergy}
\end{equation}
and the ``total energy'' is then defined as
\begin{equation}
E = \sum_i i\,N(E_i),
\label{eq:constructedE}
\end{equation}
with $N(E_i)$ the frequency (occupation) of the $i$-th ranked word-type \citep{Aerts2024Origin}.
These quantities are not measured constraints in the sense of statistical mechanics; they are constructed from the rank--frequency table by definition. For notational convenience, once \eqref{eq:rankenergy} is adopted we write $N(i)\equiv N(E_i)$. As a result, a BE-shaped fit cannot be taken as evidence for BE physics unless the mapping is operationally justified and shown to be robust.

One can also see why a BE curve can mimic familiar linguistic scaling when energy is identified with rank. With \eqref{eq:rankenergy}, the BE functional form reads
\begin{equation}
N(i)=\frac{1}{A e^{i/B}-1}.
\label{eq:bei}
\end{equation}
For $i\ll B$, $e^{i/B}=1+i/B+O((i/B)^2)$, so
\begin{equation}
N(i)\approx \frac{1}{(A-1) + (A/B)i}.
\label{eq:bei_approx}
\end{equation}
If a fit yields $A$ close to $1$, then in the same small-$i$ regime \eqref{eq:bei_approx} is approximately Zipf--Mandelbrot-like. Written in a form that avoids denominator ambiguity,
\begin{equation}
N(i)\approx \frac{B}{A\,i + B(A-1)}.
\label{eq:zipf_mandelbrot_approx}
\end{equation}
Only when the offset term is negligible, i.e.\ when
\begin{equation}
i \gg \frac{B(A-1)}{A}
\quad\text{and still}\quad i\ll B,
\label{eq:zipf_window}
\end{equation}
does \eqref{eq:zipf_mandelbrot_approx} simplify further to an approximately Zipf-like scaling:
\begin{equation}
N(i)\approx \frac{B}{i}
\qquad \text{(in the window \eqref{eq:zipf_window}, with $A\approx 1$).}
\label{eq:zipf_asympt}
\end{equation}
In other words, one gets a Zipf-like window only when $A$ is sufficiently close to $1$ and there exists an intermediate range of ranks satisfying \eqref{eq:zipf_window}. This is not a refutation of the fit; it is a reminder that, under rank-as-energy, the BE form has enough flexibility to reproduce classical rank--frequency regularities over an intermediate range \citep{Zipf1935,Mandelbrot1953}.

For completeness, note that for sufficiently large $i$ such that $A e^{i/B}\gg 1$, one has the exponential-tail approximation
\begin{equation}
N(i)\approx \frac{1}{A}\,e^{-i/B},
\label{eq:bei_tail}
\end{equation}
so in this parametrization one generically expects a \emph{soft} high-rank decay (often described informally as a ``cutoff,'' but not a sharp truncation).

\subsection{The baseline comparison is too weak for the conclusion}
The paper contrasts its BE-like fit with an exponential form labelled ``Maxwell--Boltzmann'' \citep{Aerts2024Origin}. Under the rank-as-energy identification, this baseline effectively becomes exponential decay in rank, which is not a strong competitor for heavy-tailed rank--frequency data. Consequently, the observation ``BE fits better than MB'' is not, by itself, evidence that the underlying mechanism is bosonic.
A more informative test would compare BE-shaped fits against standard linguistic and statistical baselines (e.g.\ Zipf--Mandelbrot, log-normal, stretched exponential) using the \emph{discrete likelihood} appropriate for count data, with uncertainty estimates and out-of-sample validation (e.g.\ cross-validated log-likelihood), alongside likelihood-based criteria such as AIC/BIC when appropriate \citep{Clauset2009}.

\section{A few additional concerns (brief)}
\begin{itemize}
\item \textbf{Interpretation drift:} The paper moves back and forth between ``quantum-like'' modeling language and stronger ontological claims (bosons, BE temperature, origin of quantum statistics). Those stronger claims need correspondingly stronger evidence.
\item \textbf{Parameter semantics:} Treating the fitted parameter $B$ as a literal ``temperature'' or ``heat'' is not justified when the energy axis is defined as rank and no ensemble derivation (including $\mu$, constraints, and density of states) is provided \citep{PathriaBeale}.
\item \textbf{Randomization and mechanism:} The paper reports that a BE-like fit remains good even after randomizing word order. That result weakens any causal story that assigns the BE form primarily to sentence-level meaning dynamics \citep{Aerts2024Origin}.
\item \textbf{Reproducibility:} For a quantitative claim, the work should specify the corpus, preprocessing steps (tokenization, case-folding, stopwords, lemmatization), the fitting procedure, uncertainty estimates, and the model-selection criterion.
\end{itemize}

\section{What would make the proposal genuinely testable?}
A minimal strengthening of the paper’s case would include:
\begin{enumerate}
\item \textbf{Operationalize ``energy'':} Replace $E_i=i$ with an empirically grounded cost (e.g.\ surprisal or description length) and show the conclusions are stable under reasonable alternatives.
\item \textbf{Use serious model selection:} Compare against Zipf--Mandelbrot and log-normal baselines using discrete MLE, AIC/BIC (where appropriate), and cross-validated (out-of-sample) log-likelihood \citep{Clauset2009}.
\item \textbf{Clarify the ``quantum'' content:} If terms like ``indistinguishability'' or ``entanglement'' are retained, define the analogue of state preparation and observables (and identify some nonclassical constraint beyond correlation).
\end{enumerate}

\section{Conclusion}
\citet{Aerts2024Origin} present an intriguing analogy and a visually strong fit. However, the step from ``a BE-shaped curve fits the ranked frequencies'' to ``texts behave as an ideal Bose gas'' is not supported by the current derivations and comparisons. The normalization-based ``bosonic enhancement'' argument is not a valid probability claim; rank-as-energy makes the BE fit only weakly diagnostic; and the baseline comparison is too weak to ground an ontological interpretation. Moreover, under rank-as-energy the BE form generically implies a soft exponential tail at high rank. With operational definitions and rigorous model selection, the work could be reframed as a careful phenomenology of rank--frequency curves---but not, as written, as evidence for Bose statistics in language, nor as insight into the physical origin of quantum statistics.

\end{document}